\documentclass[aps,pre,onecolumn,12pt]{revtex4}
\usepackage{amsmath}
\usepackage{dcolumn}
\usepackage{graphicx}
\usepackage{subfigure}
\usepackage{float}

\begin{document}

\title{Interactions between two-dimensional solitons in the
diffractive-diffusive Ginzburg-Landau equation with the cubic-quintic
nonlinearity}
\author{George Wainblat, Boris A. Malomed}
\affiliation{Department of Physical Electronics, School of Electrical Engineering,
Faculty of Engineering, Tel Aviv University, Tel Aviv 69978, Israel}

\begin{abstract}
We report results of systematic numerical analysis of collisions between two
and three stable dissipative solitons in the two-dimensional (2D) complex
Ginzburg-Landau equation (CGLE) with the cubic-quintic (CQ) combination of
gain and loss terms. The equation may be realized as a model of a laser
cavity which includes the spatial diffraction, together with the anomalous
group-velocity dispersion (GVD) and spectral filtering acting in the
temporal direction. Collisions between solitons are possible due to the
Galilean invariance along the spatial axis. Outcomes of the collisions are
identified by varying the GVD coefficient, $\beta $, and the collision
``velocity" (actually, it is the spatial slope of the soliton's trajectory).
At small velocities, two or three in-phase solitons merge into a single
standing one. At larger velocities, both in-phase soliton pairs and pairs of
solitons with opposite signs suffer a transition into a delocalized chaotic
state. At still larger velocities, all collisions become quasi-elastic. A
new outcome is revealed by collisions between slow solitons with opposite
signs: they self-trap into persistent \textit{wobbling dipoles}, which are
found in two modifications -- horizontal at smaller $\beta $, and vertical
if $\beta $ is larger (the horizontal ones resemble ``zigzag" bound states
of two solitons known in the 1D CGL equation of the CQ type). Collisions
between solitons with a finite mismatch between their trajectories are
studied too.
\end{abstract}

\maketitle

\section{Introduction}

Complex Ginzburg-Landau equations (CGLEs) constitute a vast class of models
for the pattern-formation dynamics and spatiotemporal chaos in one- and
multidimensional nonlinear media combining dissipative and
dispersive/diffractive properties \cite{CGL}. In particular, stable
localized pulses (``dissipative solitons" \cite{DS}) can be supported by
CGLEs that meet the obvious necessary condition of the stability of the zero
background. This condition rules out the simplest cubic CGLE, whose
one-dimensional (1D) variant admits well-known exact analytical solutions
for solitary pulses \cite{Lennart}. The stability can be achieved in systems
of linearly coupled equations, with one featuring linear gain and the other
-- linear loss \cite{coupled}. In such a model, exact stable solutions for
1D solitons are available \cite{exact}. Another possibility is to use the
CGLE with the cubic-quintic (CQ) combination of nonlinear terms. For the
first time, the CGLE of the CQ type was introduced by Petviashvili and
Sergeev \cite{PetSer} in the 2D form, with the intention to construct stable
fully localized 2D states. In 1D, stable dissipative solitons of the CQ CGLE
had been later studied in detail \cite{Thual}, including the analysis of
two-soliton bound states \cite{bound,oscill}. Then, stable fundamental
solitons \cite{Deissler,we,HS} and localized vortices (alias spiral
solitons) \cite{Bucharest,HS} have been found in 2D and 3D \cite{3D} models
of the CQ-CGLE type, as well as in the Swift-Hohenberg equation with the CQ
nonlinearity \cite{SB}. Such equations find their most significant physical
realization as models of large-area laser cavities, where the CQ combination
of the loss and gain is provided by the integration of linear amplifiers and
saturable absorbers \cite{cavity}.

In most above-mentioned works \cite{PetSer}, \cite{HS}-\cite{SB}, localized
pulses and vortices were obtained as solutions to isotropic 2D equations. On
the other hand, the CGLE which governs the spatiotemporal evolution of light
in the large-area laser cavity is anisotropic, as its includes ``diffusion"
(the spectral filtering) acting only along the temporal variable. The
existence of stable fully localized pulse solutions in the latter case
suggest a possibility of the experimental creation of ``light bullets",
i.e., spatiotemporal optical solitons, in the cavities. In other physical
contexts (unrelated to optics), anisotropy of the 2D CGLE was introduced in
a different form, through unequal diffusion coefficients in the two
perpendicular directions \cite{Rabin}.

In Refs. \cite{we}, stable spatiotemporal dissipative solitons were found in
the model of the laser-cavity type, based on the following normalized CGLE
with the CQ nonlinearity:
\begin{equation}
iU_{Z}+\frac{1}{2}U_{XX}+\frac{1}{2}\left( \beta -i\right) U_{TT}=-\left[
iU+\left( 1-i\gamma _{1}\right) |U|^{2}U+i\gamma _{2}|U|^{4}U\right] .
\label{CQfinal}
\end{equation}%
Here, $Z$ and $X$ are the propagation and transverse coordinates in the
cavity, and $T\equiv t-Z/V_{0}$ is, as usual, the reduced time, with $t$ the
physical time and $V_{0}$ the group velocity of the carrier wave. Term $%
U_{XX}$ in Eq. (\ref{CQfinal}) represents the transverse diffraction in the
paraxial approximation, the coefficients accounting for the above-mentioned
spectral filtering, Kerr nonlinearity, and background linear loss are all
scaled to be $1$, while $\beta >0$ corresponds to the group-velocity
dispersion (GVD). Usually, a necessary condition for the existence of
temporal solitons is $\beta >0$ \cite{Agr}, which implies the anomalous type
of the GVD (in the present model, spatiotemporal solitons also tend to be
more stable at $\beta >0$ \cite{we}). Further, positive coefficients $\gamma
_{1}$ and $\gamma _{2}$ in Eq. (\ref{CQfinal}) account for the cubic gain
and quintic loss, respectively, which are characteristic features of CQ
models. The third-order GVD was also taken into regard in Refs. \cite{we},
but this term is not considered here, as it does not essentially affect the
results reported below. Because it combines the diffraction along $X$ and
effective diffusion along $T$, Eq. (\ref{CQfinal}) is called the
diffractive-diffusive CGLE \cite{we}.

Once 2D solitons are available, an issue of obvious interest is to explore
collisions between them, provided that they are mobile, i.e., the equation
is Galilean invariant. The 2D CGLE with no diffusion obviously satisfies
this condition, allowing free motion of solitons or localized vortices in
any direction. This property was used in Ref. \cite{HS} to study collisions
between solitons in the isotropic CQ CGLE, as well as their motion in
external potentials. It was concluded that collisions between fundamental
solitons result in their quasi-elastic passage through each other (with a
resultant \textit{increase} of the relative velocity), or mutual destruction
of the solitons, or their merger into a single 2D pulse. In the same model,
collisions between vortices demonstrated a quasi-elastic rebound.

The laser-cavity model based on Eq. (\ref{CQfinal}) features the Galilean
invariance along the $X$-direction, which means that a moving solution can
be generated from a quiescent one by the application of the Galilean boost
corresponding to arbitrary ``velocity" $P$ (in fact, $P$ is the tilt in the $%
\left( X,Z\right) $ plane):
\begin{equation}
U\left( X,T,Z\right) \rightarrow \exp \left[ i\left( PX-P^{2}Z/2\right) %
\right] U\left( X-PZ,T,Z\right) .  \label{boost}
\end{equation}%
This possibility suggests to consider collisions of 2D solitons in this
model too. In this work, we report results obtained by means of systematic
simulations of collisions between two and three solitons in the framework of
Eq. (\ref{CQfinal}). In the former case, both head-on collisions and those
with a finite offset (\textit{aiming distance}) between trajectories of the
two solitons will be studied. In either case, we consider collisions between
in-phase and out-of-phase 2D solitons (the latter means that they have
opposite signs).

In Section II, we report the results for two-soliton collisions, and in
Section III -- for interactions between three solitons. Outcomes of the
collisions between two in-phase solitons include the quasi-elastic passage
at large velocities, delocalization in the $X$-direction (merger into an
expanding quasi-turbulent state) at intermediate velocities, and merger of
slowly moving solitons into a single stable pulse. A major difference for
collisions between out-of-phase solitons is that, at small velocities, they
do not merge into a single pulse; instead, they may form a new localized
object -- a \textit{wobbling dipole}, i.e., a robust bound state of two
solitons with opposite signs, which feature persistent oscillations relative
to each other in the spatial direction. Moreover, two different species of
the wobbling dipoles are reported below, \textit{horizontal} and \textit{%
vertical} ones. In the latter case, the out-of-phase solitons, although they
collide head-on, shift in opposite perpendicular directions\ (along the $T$%
-axis), and eventually form a dipole with a fixed vertical separation
between them. Unlike the results for collisions between dissipative solitons
in the 2D isotropic CGLE \cite{HS}, in the present model we have never
observed complete destruction (decay) of colliding solitons. For collisions
with a finite aiming distance $\Delta T$, we identify a critical value of $%
\Delta T$ which separates interactions and the straightforward passage.
Three-soliton configurations feature either merger into a single pulse, or
the transition into a delocalized chaotic state. In terms of the optical
cavities, the various outcomes of the collisions offer possibilities for the
use in all-optical data-processing schemes.

\section{Two-soliton collisions}

\subsection{The numerical procedure}

Equation (\ref{CQfinal}) was solved by means of the 2D split-step Fourier
method with $256\times 256$ modes and periodic boundary conditions in $X$
and $T$, for the fixed size of the integration domain in both directions, $%
\left\vert X,T-10\right\vert \leq 10$. The stepsize for the advancement in $%
Z $ was $0.01$. To generate the first stable 2D pulse boosted to velocity
(tilt) $P$, cf. Eq. (\ref{boost}), an initial configuration was taken as
\begin{equation}
U_{0}\left( X,T\right) =\exp \left[ -\left( X^{2}+T^{2}\right) /4+iPX\right]
,  \label{initial}
\end{equation}%
see the first panel in Fig. \ref{fig2} below. The numerical integration of
Eq. (\ref{CQfinal}) led to quick self-trapping of the input pulse into a
moving (tilted) dissipative soliton, which is an attractor of the model. The
profile of the established soliton can be seen in the first panels of Figs. %
\ref{fig3} and \ref{fig5}.

Generic results for collisions between the solitons with velocities $\pm P$
can be adequately represented by fixing the cubic gain and quintic loss
coefficients to be $\gamma _{1}=2.5$, $\gamma _{2}=0.5$, while varying $P$
and GVD coefficient $\beta $. To generate diagrams presented below in Figs. %
\ref{fig1} and \ref{fig4}, which display outcomes of the collisions, we
changed $\beta $ and $P$ by small steps, the initial configuration for each
simulation being a stable pulse produced by the simulations at the previous
step.

\subsection{Head-one collisions between in-phase solitons}

\begin{figure}[tbp]
{\includegraphics[width=.8\linewidth]{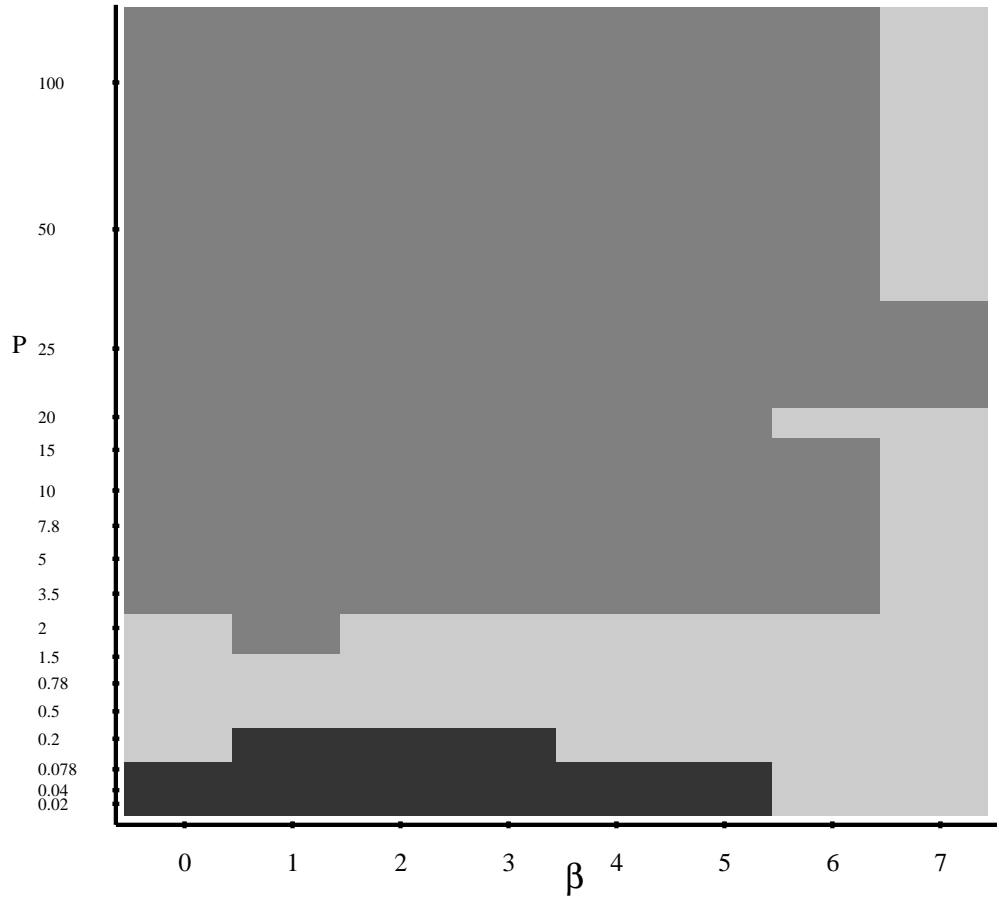}}
\caption{The diagram of outcomes of head-on collisions between identical
in-phase solitons, moving at velocities $\pm P$ (the values of $P$ are shown
on the logarithmic scale). The black, light gray, and dark gray colors mark
parameter regions where the merger, transition to a chaotic delocalized
state, and quasi-elastic passage have been observed.}
\label{fig1}
\end{figure}

Outcomes of collisions between two identical stable solitons, set by kicks $%
\pm P$ on the head-on collision course, are summarized in Fig. \ref{fig1}.
Stable solitons exist only for $\beta \geq \beta _{\min }\approx -0.5$,
which determines the left-hand edge of the diagram.

The simplest outcome of the collision is the straightforward quasi-elastic
passage of the solitons through each other. We do not illustrate it by a
separate picture, as it seems quite obvious; as well as in Refs. \cite{we},
the solitons keep the mutual symmetry after the quasi-elastic collision, and
demonstrate some increase of ``velocity" $P$ (recall it is actually defined
as the tilt of the soliton's trajectory in the $\left( X,Z\right) $ plane).
Moreover, running the simulations in the domain with periodic boundary
conditions, we observed multiple quasi-elastic collisions between solitons.
The solitons which emerge unscathed from the first collision survive
indefinitely many repeated collisions as well.

\begin{figure}[tbp]
\includegraphics[width=.8\linewidth]{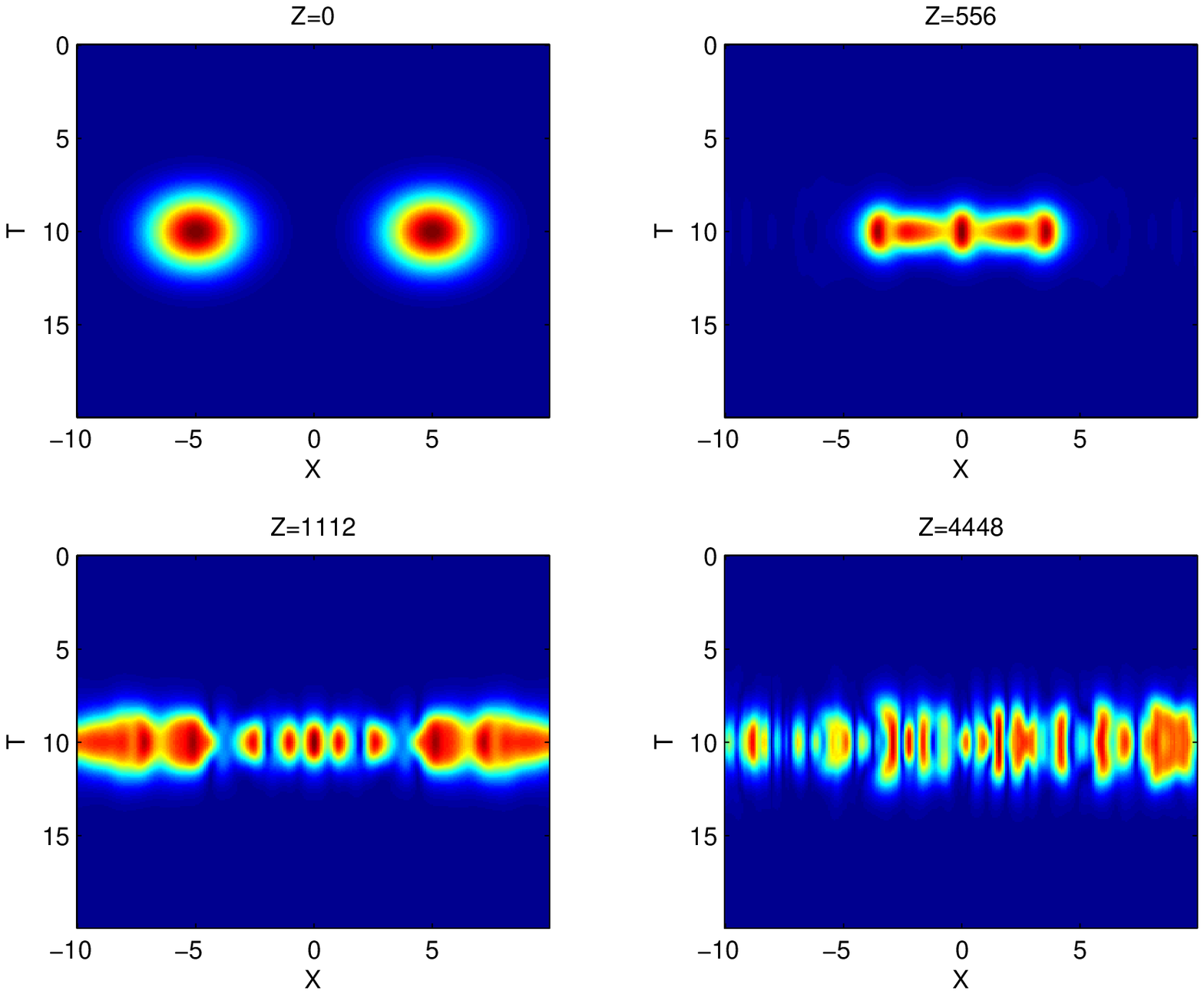}
\caption{(Color online) A typical example of the transition to
delocalization along the $x$ direction, triggered by the merger of colliding
in-phase solitons with velocities $P=\pm 0.78$, for $\protect\beta =1$. The
evolution of the wave field is illustrated by a set of snapshots of the
distribution of $\left\vert U\left( X,T\right) \right\vert $. Note that the
profiles displayed in the first panel (for $Z=0$) pertain not to the
established solitons, but to input (\protect\ref{boost}). Profiles of the
established solitons generated from these inputs can be seen below in the
first panels of Figs. \protect\ref{fig3} and \protect\ref{fig5}.}
\label{fig2}
\end{figure}

With the decrease of the collision velocity, the quasi-elastic passage is
changed by the delocalization. This means that two in-phase solitons,
interacting attractively, merge into a single pulse, which, however, fails
to self-trap into a standing soliton. Instead, it gives rise to a
quasi-chaotic (``turbulent") state, that remains localized in the temporal
direction ($T$), but features indefinite expansion along $X$, see a typical
example in Fig. \ref{fig2}. This outcome may be explained by the fact the
fused state has too much ``intrinsic inertia", imparted by original
velocities $\pm P$, which pushes the pulse to expand.

\begin{figure}[tbp]
\includegraphics[width=.8\linewidth]{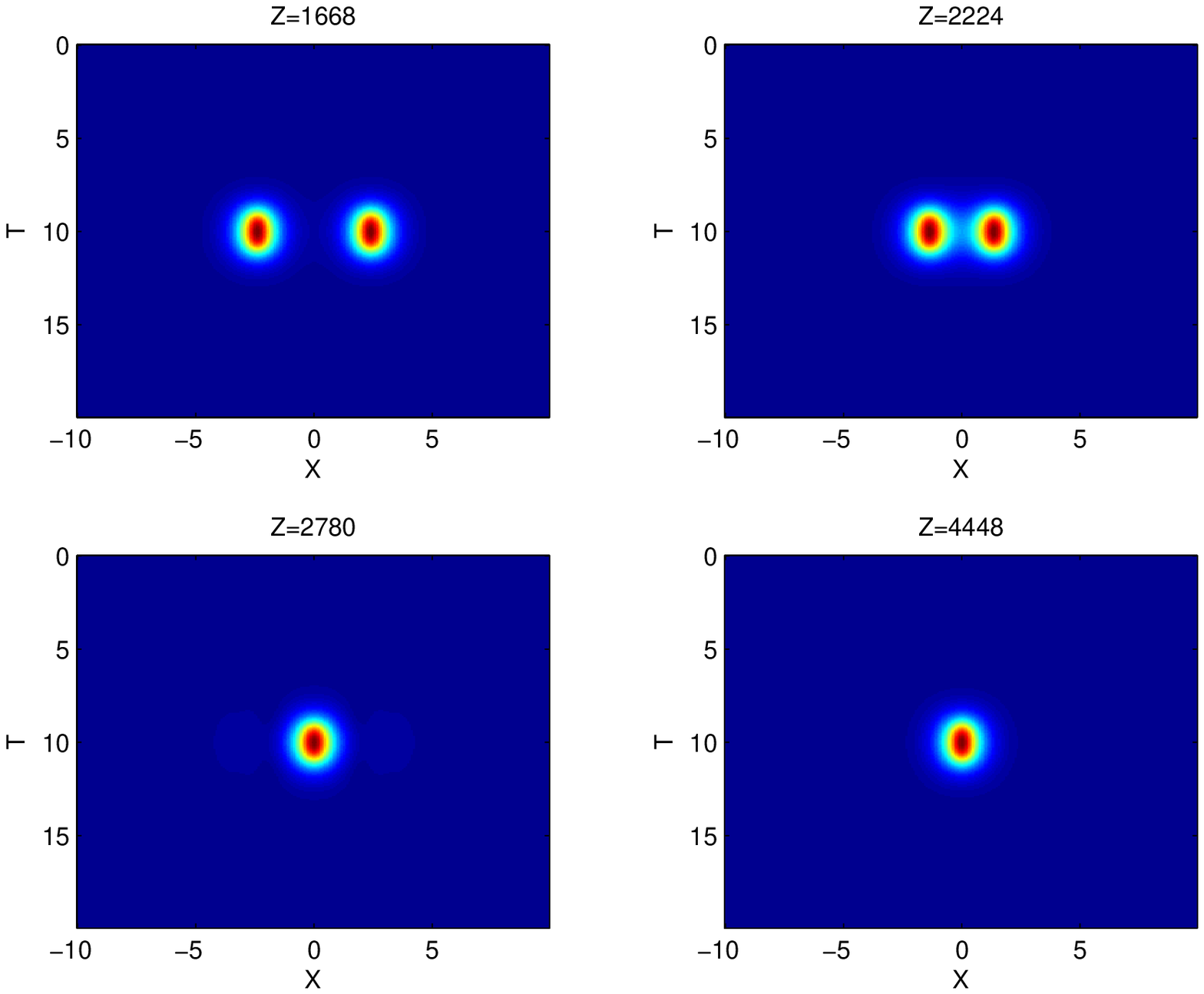}
\caption{(Color online) An example of the merger of colliding in-phase
solitons into a single stable one, for $P=\pm 0.078$, $\protect\beta =1$.}
\label{fig3}
\end{figure}

At still smaller values of $P$, the collision also gives rise to merger of
the two solitons into a single pulse. However, in that case, the decrease of
the above-mentioned ``intrinsic inertia" allows the fused pulse to form a
stable soliton, see a typical example in Fig. \ref{fig3}.

Before proceeding to the presentation of results obtained for collisions
between solitons with opposite signs, it is relevant to mention that we have
also considered collisions of solitons with the phase difference of $\pi /2$%
. In that case (not shown here in detail), the merger of two solitons into a
single one is also observed at small velocities, but under more specific
conditions. In particular, for $P\leq 0.78$, the merger takes place in the
region of $1\leq \beta \leq 3$, which is essentially narrower than the
merger region for in-phase solitons, cf. Fig. \ref{fig1}. The reduction of
the merger region is quite natural, as the interaction between the solitons
is weaker in this case than between in-phase solitons.

\subsection{Head-on collisions between solitons with opposite signs}

\begin{figure}[tbp]
{\includegraphics[width=.8\linewidth]{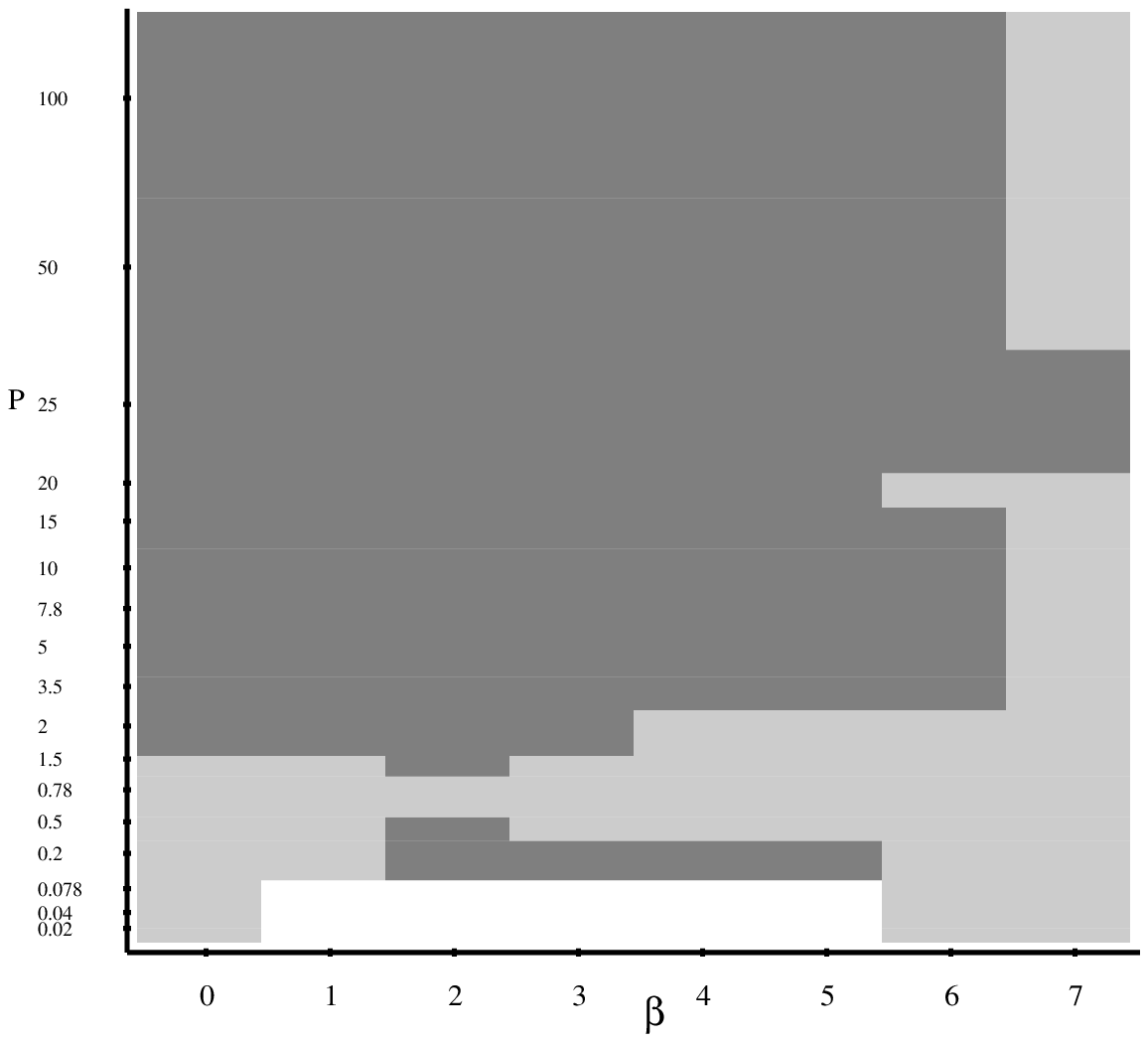}}
\caption{The diagram of head-on collisions between identical solitons with
opposite signs. As in Fig. \protect\ref{fig1}, the light and dark gray
colors designate regions of the transition to chaotic delocalized states,
and quasi-elastic passage, respectively. In the white area, the slowly
colliding solitons form horizontal or vertical \textit{wobbling dipoles},
see the text.}
\label{fig4}
\end{figure}

The diagram summarizing outcomes of collision between solitons with opposite
signs is displayed in Fig. \ref{fig4}.
The out-of-phase solitons with large velocities $\pm P$ pass through each
other quasi-elastically. Moreover, because the simulations were run with the
periodic boundary conditions, we could actually observe multiple collisions,
which kept their quasi-elastic character indefinitely long, similar to what
was observed in the case of collisions between fast in-phase solitons, as
mentioned above. At smaller (intermediate) velocities, collisions between
the solitons with opposite signs give rise to the formation of chaotic
states indefinitely expanding along $X$, again similar to what was reported
above for the case of in-phase solitons (therefore, examples of these
outcomes are not displayed here).

\begin{figure}[tbp]
$%
\begin{array}{cc}
\includegraphics[width=.52\linewidth]{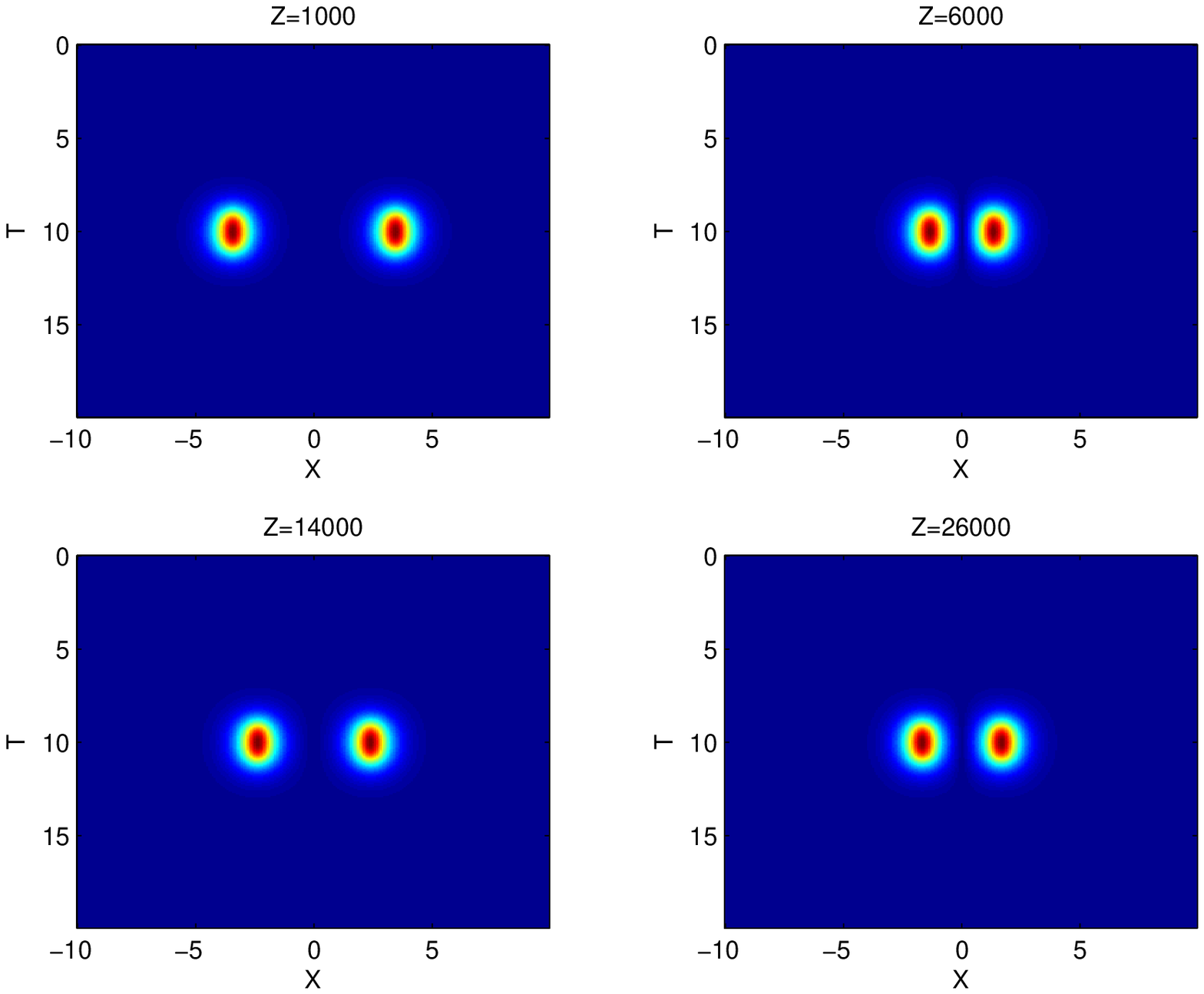} & \includegraphics[width=.48%
\linewidth]{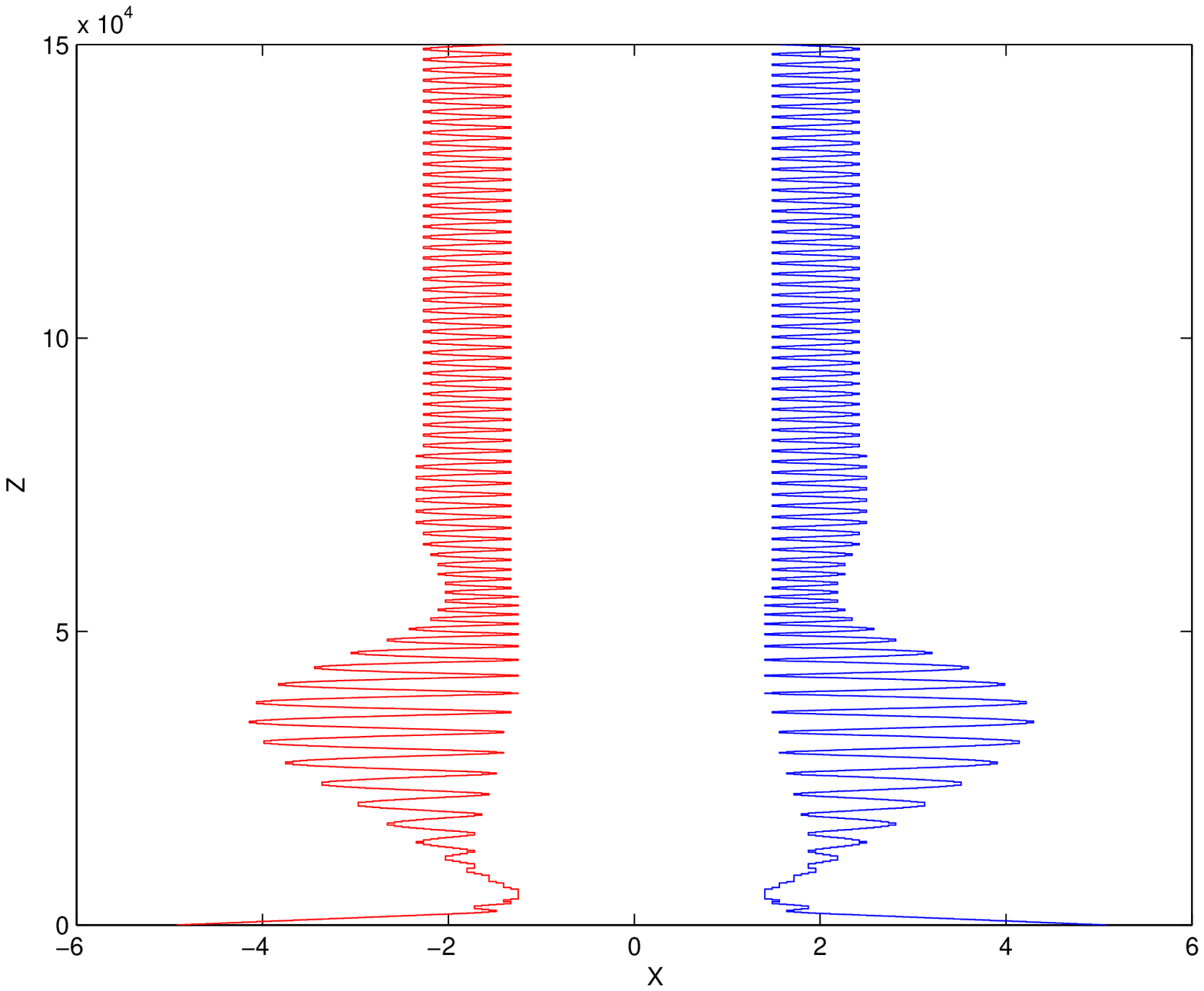}%
\end{array}%
$%
\caption{(Color online) Panels in the left-hand half illustrate the
formation of a horizontal ``wobbling dipole" as a result of the head-on
collision between slowly moving out-of-phase solitons, for $P=\pm 0.078$, $%
\protect\beta =1$. The panel in the right-hand half shows trajectories of
centers of the two solitons, in the same case.}
\label{fig5}
\end{figure}

A new outcome is produced by collisions of slowly moving out-of-phase
solitons. They do not merge into a single pulse, because this is prevented
by the mutual repulsion. As seen in the left-hand part of Fig. \ref{fig5},
the solitons come close to each other and then bounce back, but do not
separate. Instead, they arrange themselves into a persistent bound state in
the form of a ``wobbling dipole". This observation is illustrated by the
right-hand side of Fig. \ref{fig5}, which displays trajectories of centers
of both solitons in the plane of $\left( X,Z\right) $. Persistent
oscillations of the bound solitons lasted as long as the simulations were
run. Note that the range of the evolution distance in Fig. \ref{fig5}, $%
Z=15\cdot 10^{4}$, is extremely large in comparison with the soliton's
diffraction, dispersion, and filtering lengths, which can be estimated as $%
Z_{\mathrm{diffr}}\sim Z_{\mathrm{disp}}\sim Z_{\mathrm{filt}}\sim 1$ in the
present situation, following the usual definitions, $Z_{\mathrm{diffr}}\sim
W_{X}^{2}$, $Z_{\mathrm{disp}}\sim W_{T}^{2}/\beta $, $Z_{\mathrm{filt}}\sim
W_{T}^{2}$, where $W_{X}$ and $W_{T}$ are the soliton's widths in the
spatial and temporal directions \cite{Agr} (here, the notation is implied to
be as in Eq. (\ref{CQfinal})). Dependences of the amplitude and frequency of
the oscillations on GVD coefficient $\beta $ is not shown here, as the
dependence is quite weak.

It is relevant to mention that similar stable wobbling bound states of
dissipative solitons were reported in simulations of a system of two CQ
CGLEs coupled by cubic terms, in the 1D setting, with a group-velocity
mismatch between the two equations \cite{Brand} (in those works, they were
called ``zigzag" states). On the other hand, in the framework of the single
CQ CGLE in 1D, truly stable bound states do not exist, although some of them
may be almost stable \cite{oscill}, with the difference that the phase shift
between the two solitons is $\pi /2$ (rather than $\pi $). Thus, the
existence of the stable oscillatory dipolar bound states in the single
equation is a new feature of the 2D setting.

\begin{figure}[tbp]
$%
\begin{array}{cc}
\includegraphics[width=.52\linewidth]{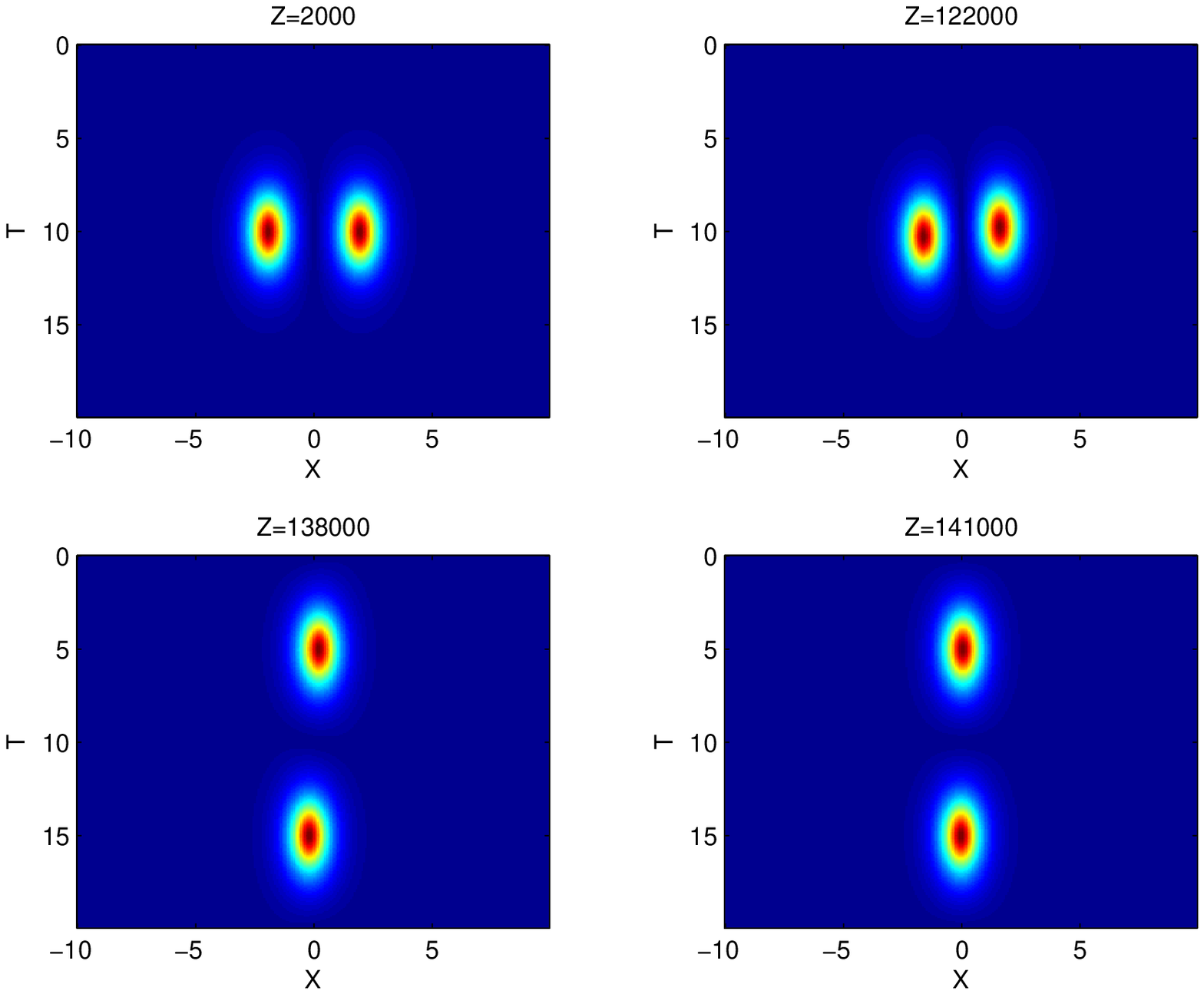} & \includegraphics[width=.48%
\linewidth]{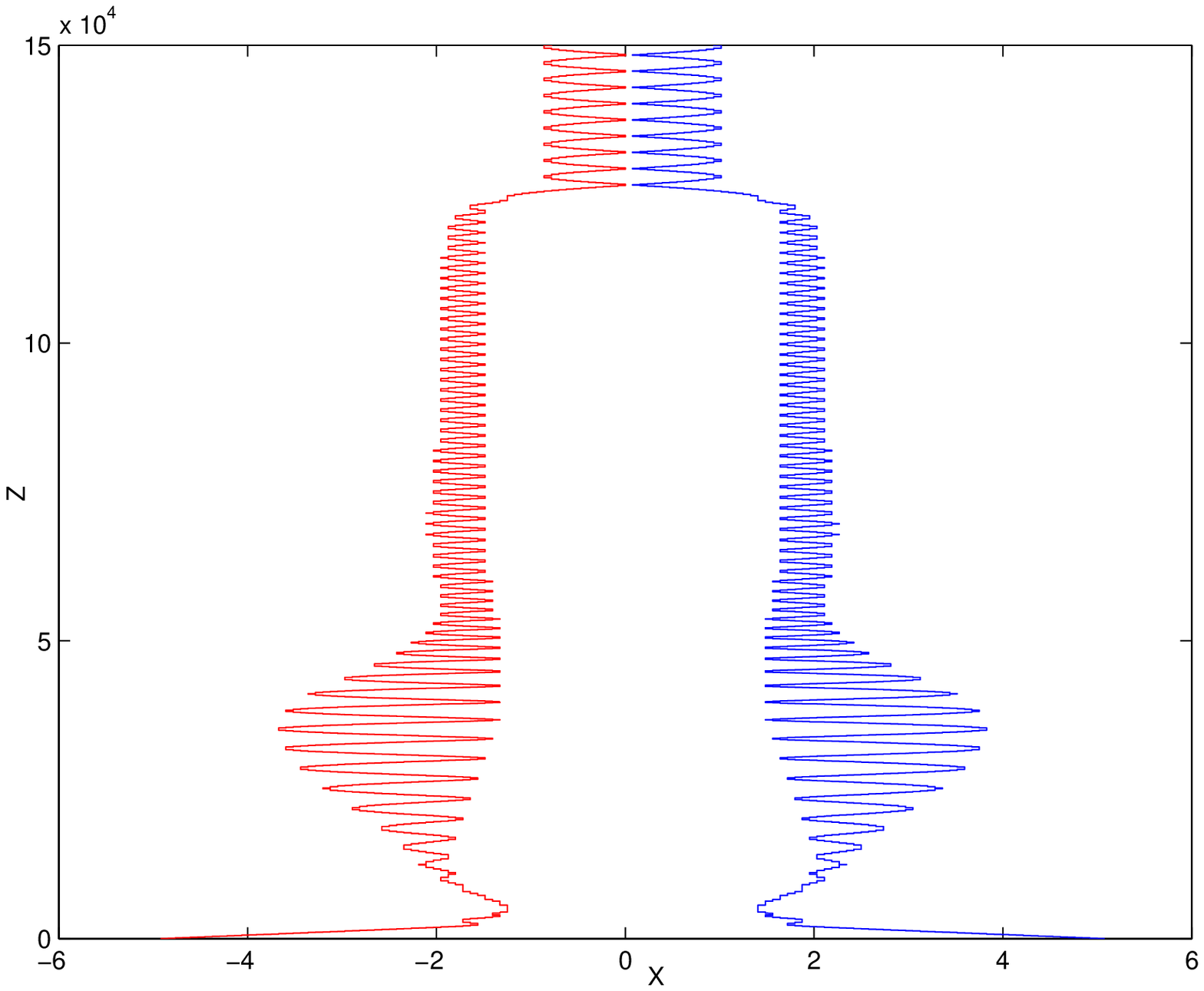}%
\end{array}%
$%
\caption{(Color online) The same as in Fig. \protect\ref{fig5}, but in the
case when the collision of slowly moving solitons with opposite signs ends
up with the formation of the vertical ``wobbling dipole", at $\protect\beta %
=5$.}
\label{fig6}
\end{figure}

The formation of the wobbling dipole follows the above scenario in the
region of $\beta \leq 3$. At larger values of the GVD coefficient (in
particular, at $\beta =4$ and $5$), the repulsive interaction of slowly
moving out-of-phase solitons leads to their shift in the transverse
direction (to positive and negative values of $T$). Eventually, they form a
vertically aligned (``stacked") wobbling dipole, as illustrated by a set of
snapshots in the left-hand side of Fig. \ref{fig6}. The vertical dipole
keeps a constant vertical (i.e., temporal) separation between the two
solitons stacked in it, which simultaneously perform persistent oscillations
in the horizontal direction (along $X$), as seen in the right-hand side of
Fig. \ref{fig6}.

\subsection{Collisions at a finite aiming distance}

Collisions between slowly moving in-phase solitons, with a finite offset, $%
\Delta T$, between their initial trajectories in the vertical direction
(alias the aiming distance), also result in the merger into a single
quiescent pulse, provided that both $\Delta T$ and velocities are small
enough. However, the transition to the delocalization, similar to that shown
in Fig. \ref{fig2}, was not observed at finite $\Delta T$. In the same case,
collisions between solitons with opposite signs result in a simple dynamical
effect, a rebound in the vertical direction. Namely, due to the repulsion
between the out-of-phase solitons, the value of $\Delta T$ increases after
the collision. These outcome are not shown here, as they are quite obvious.

\begin{figure}[tbp]
{\includegraphics[width=.8\linewidth]{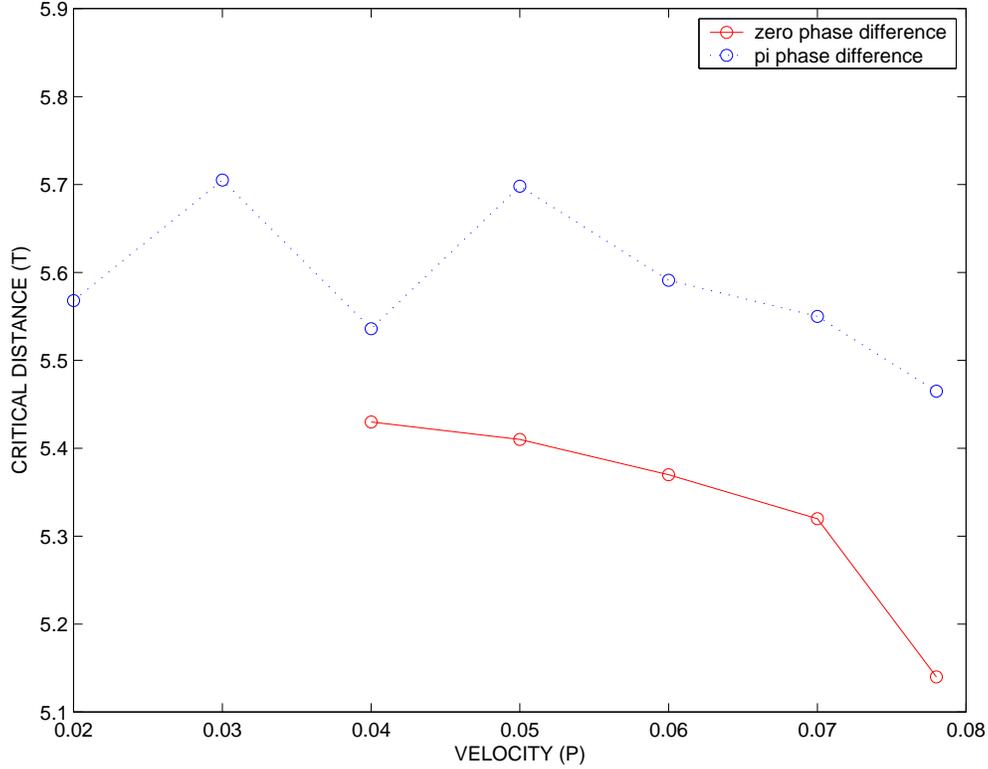}}
\caption{(Color online) The critical value of the offset between initial
trajectories of two solitons ($\Delta T$ in the text), in the case of
non-head-on collisions. If the offset exceeds the critical value, the
solitons effectively cease to interact. This value is shown versus the
initial velocities, $\pm P$, separately for in-phase and out-of-phase
soliton pairs.}
\label{fig8}
\end{figure}

Varying collision velocities $\pm P$, it is possible to find a critical
value of the offset, such that the interaction becomes negligible if $\Delta
T$ exceeds the critical offset. For both cases of the in-phase and $\pi $%
-out-of phase soliton pairs, the critical value is shown, as a function of
the velocity, in Fig. \ref{fig8}.

\section{Three-solitons collisions}

Once the character of the two-soliton collisions has been understood, the
next natural step is to analyze collisions between three solitons. To this
end, we took triplets of identical solitons, with initial velocities $+P,$ $%
0,$ $-P$. The systematic analysis was restricted to the most interesting
case of the collisions between slow solitons, with $P\leq 0.078$.

\begin{figure}[tbp]
\includegraphics[width=.8\linewidth]{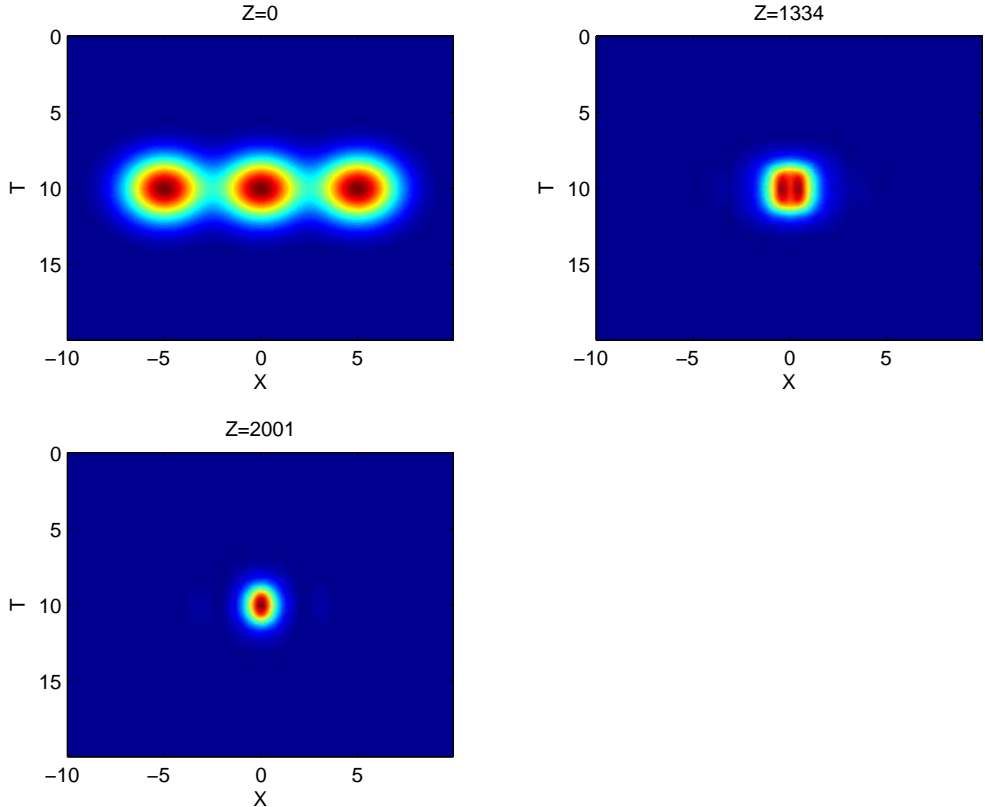}
\caption{(Color online) The merger of three in-phase solitons into a single
one, at $\protect\beta =1$. Initially, edge solitons move at velocities $\pm
0.078$.}
\label{fig9}
\end{figure}

First, we consider symmetric configurations, with solitons' signs
$\left( +,+,+\right) $ and $\left( +,-,+\right) $. In the former
case, the outcome of the collision in the entire region where the
simulations were run, $0\leq \beta \leq 7$, is the merger of the
triplet of in-phase solitons into a single pulse, see an example in
Fig. \ref{fig9}. In the latter case, the triplet of solitons with
alternating signs always features the transition to delocalization.
A noteworthy peculiarity observed in the case of $\left(
+,-,+\right) $ is formation of a transient ``wobbling tri-pole"
configuration, that qualitatively resembles horizontal wobbling
dipoles generated by collisions of two solitons with opposite signs,
see Fig. \ref{fig5}. Nevertheless, the ``tri-pole" eventually
collapses, initiating the transition into an expanding chaotic
delocalized state.

\begin{figure}[tbp]
\includegraphics[width=.8\linewidth]{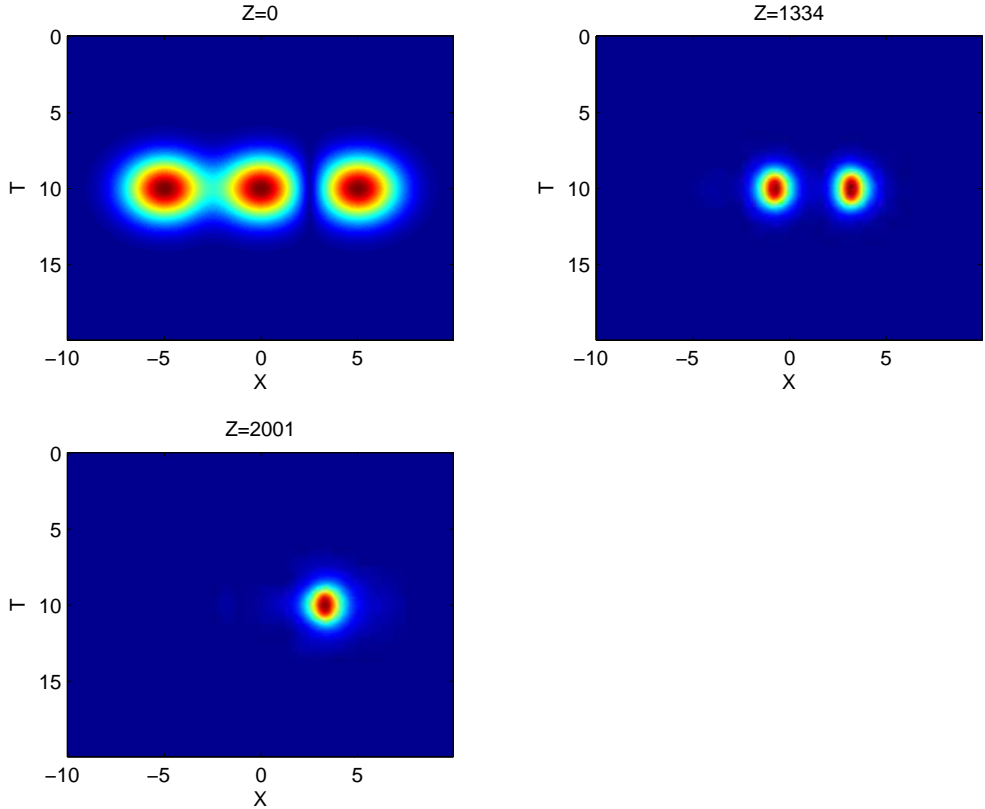}
\caption{(Color online) The same as in Fig. \protect\ref{fig9}, but starting
with the asymmetric triplet of solitons, configured as $\left( +,+,-\right) $%
.}
\label{fig11}
\end{figure}

The collision-induced transformation of the asymmetric triplet formed by
three identical solitons, of type $\left( +,+,-\right) $, was investigated
too. As shown in Fig. \ref{fig11}, in this case, two in-phase solitons merge
into a single one, at the first stage of the evolution (the middle panel in
Fig. \ref{fig11}). Eventually, the intermediate pair of the two pulses also
demonstrates a transformation into a single residual soliton.

\section{Conclusion}

We have undertaken the systematic analysis of collisions between two and
three solitons in the 2D CGLE (complex Ginzburg-Landau equation) with the
CQ\ (cubic-quintic) nonlinearity, which may be considered as a model of
large-area laser cavities, with the solitons representing spatiotemporal
``light bullets" in it. This model, which includes the diffraction in the
spatial direction ($X$) and both the GVD (group-velocity dispersion) and
spectral filtering in the temporal direction, is Galilean invariant in the
former direction, which makes it possible to create moving solitons and
collide them. Outcomes of the collisions were systematically studied by
varying the GVD coefficient, $\beta $, collision velocity, and relative sign
of the solitons.

In the case of collisions between two in-phase solitons, three outcomes have
been identified: the merger into a single standing soliton, transition into
an expanding delocalized chaotic state, and quasi-elastic passage.
Collisions between solitons with opposite signs at small velocities lead,
instead of the merger, to a novel outcome -- the formation of robust
``wobbling dipoles", which exist in two modifications: horizontal at smaller
values of $\beta $, and vertical ones at larger $\beta $. In the 1D setting,
oscillatory bound states of dissipative solitons of the ``zigzag" type were
found in a system of two nonlinearly coupled CQ CGLEs \cite{Brand}, but
truly stable bound states in the single equation of this type do not exist.
The stable ``wobbling dipoles", especially their ``vertical" variety (the
one shown in Fig. \ref{fig6}), is a feature specific to the 2D model. We
have also investigated two-soliton collisions with a finite aiming distance,
and identified its critical size beyond which the solitons cease to interact.

Collisions between three slowly moving in-phase solitons lead to their
merger into a single one, while three solitons with alternating signs form a
transient ``tri-pole" state, which eventually collapses into a delocalized
chaotic state. Three solitons which form an asymmetric configuration, of
type $\left( +,+,-\right) $, also merge into a single pulse.

\end{document}